# From vectors to mnesors

Gilles CHAMPENOIS

Collège Saint-André, Saint-Maur, France

gilles_champenois@yahoo.fr

ABSTRACT. The mnesor theory is the adaptation of vectors to artificial intelligence. The scalar field is replaced by a lattice. Addition becomes idempotent and multiplication is interpreted as a selection operation. We also show that mnesors can be the foundation for a linear calculus.

> "I am still not satisfy with algebra, because it does not give the shortest methods or the most beautiful constructions in geometry. This is why I believe that, so far as geometry is concerned, we need still another analysis which is distinctly geometrical or linear and which will express situation directly as algebra expresses magnitude directly."
>
> *Gottfried Leibniz (1646-1706)*

## I. INTRODUCTION

Geometrical and physical quantities require the notions of point, direction and magnitude and also appropriate operations like addition and scalar multiplication [1]. The former is able to compose vectors or points, the latter to modify magnitude.

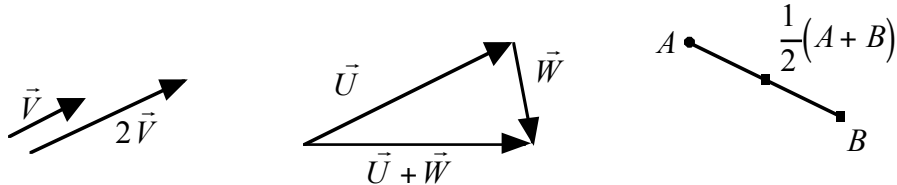

FIGURE 1. Addition and scalar multiplication of vectors or points

The axiomatization of abstract vectors [2] comes up as follows: let be given a vector space $(M,+)$ over the field $K$. $M$ has 0 as identity element and $(K,+,\cdot)$ has 1 as identity element for multiplication. For any vectors $x, y$, left module properties hold:

| | |
|---|---|
| (unital property) | $1x = x$ |
| (vector distributivity) | $\lambda(x+y) = \lambda x + \lambda y$ |
| (associativity of scalar multiplication) | $\mu(\lambda x) = (\mu \cdot \lambda)x$ |
| (scalar distributivity) | $(\lambda + \mu)x = \lambda x + \mu x$ |

But ias far as information is concerned, we'd better talk about precision instead of magnitude and about ordering instead of direction. Hence, addition should be defined as an aggregation operation and scalar multiplication as a filtering operator. For that, scalar field will be replaced by a lattice.

EXAMPLE. As simple example of mnesors, we give the column tuple $\begin{bmatrix} France \\ Poland \end{bmatrix}$. Two column tuples are added as follows:

$$\begin{bmatrix} Germany \\ Luxemburg \end{bmatrix} + \begin{bmatrix} France \\ Poland \end{bmatrix} = \begin{bmatrix} Germany \\ Luxemburg \\ France \\ Poland \end{bmatrix}$$

If $EU$ (for European Union) represents an element of the lattice of intergovernmental organizations, multiplication by $EU$ acts on a column tuple like a filter:

$$\begin{bmatrix} France \\ Russia \\ Sweden \end{bmatrix} EU = \begin{bmatrix} France \\ Sweden \end{bmatrix}$$

Only components belonging to the European Union are retained.

Because information structures and processes don't necessarily commute, we don't assume commutativity for mnesors.

EXAMPLE: $\begin{bmatrix} Slowenia \end{bmatrix} + \begin{bmatrix} Slowakia \end{bmatrix} = \begin{bmatrix} Slowenia \\ Slowakia \end{bmatrix}$ differs from $\begin{bmatrix} Slowakia \end{bmatrix} + \begin{bmatrix} Slowenia \end{bmatrix} = \begin{bmatrix} Slowakia \\ Slowenia \end{bmatrix}$

*Slowenia* comes first in the former colomn list but second in the latter. The addition is not necessarily commutative.

## II. DEFINITIONS

We define a mnesor space as a two-sorted structure made up with a monoid $(M,+)$ and a lattice $(L,\oplus,\otimes)$. We assume

    ☑ an identity element for $M$: $0$     (1)
    ☑ a top element for $L$: $\tau$     (2)

The lattice here plays the same role as a scalar field and for that reason the elements are called granular. Scalar multiplication is replaced by the *granular multiplication*, which multiplies a mnesor by any lattice element and returns another mnesor. Granular multiplication is considered as a filtering operation. The definition properties follow next:

- (unital property) $\quad x\tau = x \quad (3)$
- (mnesor distributivity) $\quad (x+y)\lambda = x\lambda + y\lambda \quad (4)$
- (associativity of granular multiplication) $\quad (x\lambda)\mu = x(\lambda \otimes \mu) \quad (5)$
- (granular distributivity) $\quad x\lambda + x\mu = x(\lambda \oplus \mu) \quad (6)$

EXAMPLE. To illustrate the unital property, you can see that $[Germany]UN = [Germany]$ where $UN$ (United Nations) is the top element of the lattice. Multiplying by the top element has no action on a mnesor.

A common assumption with the vector theory is to make addition reversible. We mean that there exists a reverse operation undoing $x + y$ and calculating $x$ from $z = x + y$. In the vector theory this operation is the substraction ($x = z - y$). We here postulate that there exists a granular doing it ($x = z\alpha = (x+y)\alpha$). Finally:

- (absorption property) For any mnesors $x, y$, there exists a granular $\alpha$ such that
$$(x+y)\alpha = x \quad (7)$$

EXAMPLE: We can now extract the term $[Italy]$ from the sum $[Italy] + [Switzerland]$ using the previous property $([Italy] + [Switzerland])NATO = [Italy]$.

## III. PROPERTIES OF NON-REDUNDANCY

*Idempotence.* The addition of mnesors is idempotent.

PROOF: Applying (4) with $\lambda = \tau$ and $\mu = \tau$ gives $x\tau + x\tau = x\tau$
Thus, $x + x = x$, for any $x \in M$ [by (3)]

EXAMPLE. $[Spain] + [Spain] = [Spain]$

*Priority.* $\quad x + y + x = x + y$, for any mnesors $x, y$

PROOF:
$x + y + x = x + y + (x+y)\alpha$ [by (7)]
$\quad = (x+y)\tau + (x+y)\alpha$ [by (3)]
$\quad = (x+y)(\tau \oplus \alpha)$ [by (6)]
$\quad = (x+y)\tau$ [by (2)]
$\quad = x + y$ [by (3)]

EXAMPLE. $\begin{bmatrix} Denmark \\ Norway \end{bmatrix} + [Denmark] = \begin{bmatrix} Denmark \\ Norway \end{bmatrix}$.

## IV. PREFIX ORDERING

If a mnesor $a$ takes the form $a = x + y$, then $x$ (resp. $y$) is a prefix (resp. suffix) for $a$.

EXAMPLE. In $\begin{bmatrix} Rumania \\ Serbia \end{bmatrix} = [Rumania] + [Serbia]$, the former is a prefix and the latter a suffix.

*Prefix.*   The next three propositions are equivalent :

(i)   $y$ is a prefix for $x$
(ii)  $y$ takes the form $y = x\lambda$
(iii) $y + x = x$

PROOF :   (i)⇒(ii)   Let $y$ be a prefix for $x$. That is, $y + z = x$.
There exists a granular $\lambda$ such that
$(y + z)\lambda = y$, or equivalently, $x\lambda = y$

(ii)⇒(iii)   from $y = x\lambda$ follows $y + x = x\lambda + x$
$$\begin{aligned} &= x\lambda + x\tau && [\text{by (3)}] \\ &= x(\lambda \oplus \tau) && [\text{by (6)}] \\ &= x\tau && [\text{by (2)}] \\ &= x && [\text{by (3)}] \end{aligned}$$

(iii)⇒(i)   evidently

*Prefix relation and ordering.*  The relation "$x$ is a prefix for $y$" give rise to an ordering.

PROOF:
- Each mnesor is a prefix for itself
- If $x$ is a prefix for $y$ (i.e. $x + y = y$) and $y$ a prefix for $z$ (i.e. $y + z = z$), then $x$ is a prefix for $z$ ($x + z = x + (y + z) = (x + y) + z = y + z = z$)
- Antisymmetry: let $x$ be a prefix for $y$ (i.e. $x + y = y$) and vice-versa (i.e. $y + x = x$), then $x + y = (y + x) + y = y + x + y$
$$\begin{aligned} &= y + x && [\text{by priority}] \\ &= x \end{aligned}$$
On the other hand $x + y = y$. Hence, $x = y$

EXAMPLE: $\begin{bmatrix} France \\ Italy \end{bmatrix} = \begin{bmatrix} France \\ Russia \\ Italy \end{bmatrix} EU$. Thus, $\begin{bmatrix} France \\ Russia \\ Italy \end{bmatrix} \geq \begin{bmatrix} France \\ Italy \end{bmatrix}$

As a result of a filtering operation, the column list gains in selectivity. The prefix ordering plays the role of a selectivity indicator.

The order is compatible with the addition and the granular multiplication.

PROOF:
- $x \leq y$ (i.e. $x + y = y$). If we add $a$ to both members, $x + y + a = y + a$, and insert $a$ then $x + a + y + a = y + a$, or equivalently $x + a \leq y + a$.

- $x \le y$ (i.e. $x + y = y$). By multiplying both members by $\lambda$ we get $x\lambda + y\lambda = y\lambda$. Thus $x\lambda \le y\lambda$.
- Let $\lambda \le \mu$ be given. That is, $\lambda + \mu = \mu$. By multiplying both members by $x$ and applying (6), we get $x\lambda + x\mu = x\mu$. Thus $x\lambda \le x\mu$.

*Positivity.* For any mnesor $x$, $x \ge 0$.

PROOF: $0 + x = x$ [by (1)]

$M$ is zerosumfree. That is, $x + y = 0$ implies $y = x = 0$

PROOF: $x$ is a prefix for $0$, that is $x + 0 = 0$
Then $x = 0$, [by (1)]
Finally, $y = 0$ [by (1)]

## V. SUFFIXES

*Suffix.* The next three propositions are equivalent :

(i)   $y$ is a suffix for $a$
(ii)  $y$ satisfies an identity of the form: $a\lambda + y = a$
(iii) $a + y = a$

PROOF : (i)$\Rightarrow$(ii)  $y$ is a suffix for $a$. That is, there exists a prefix $x = a\lambda$ such that $x + y = a$. Thus, $a\lambda + y = a$
(ii)$\Rightarrow$(iii)  Adding $a$ to both sides of $a\lambda + y = a$ gives $a + a\lambda + y = a + a$. Thus,
$a + a\lambda + y = a$ [by idempotence]
$a\tau + a\lambda + y = a$ [by (3)]
$a(\tau \oplus \lambda) + y = a$ [by (6)]
$a\tau + y = a$ [by (2)]
$a + y = a$ [by (3)]
(iii)$\Rightarrow$(i)  evidently

Any prefix for $x$ is a suffix for $x$. Hence, prefixes inherit properties from suffixes.

PROOF : Let $z$ be a prefix for $a$. That is, $z$ takes the form $z = a\lambda$
Then, $a + z = a + a\lambda$
$= a\tau + a\lambda$ [by (3)]
$= a(\tau \oplus \lambda)$ [by (6)]
$= a\tau$ [by (2)]
$= a$ [by (3)]
Thus $z$ is a suffix.

*Anagram.* Two mnesors $x, y$ are called *anagrams* if $x$ is a suffix for $y$ and vice versa. For example, $x = z + t$ and $y = t + z$ are anagrams, for any mnesors $z, t \in M$ (to prove it: $x + y = z + t + t + z = z + t = x$ and $y + x = t + z + z + z = t + z = y$).

## VI. STABILIZERS AND ANNIHLATORS

*Stabilizer*. A granular $\lambda$ is a stabilizer for a given mnesor $x$ if multiplying by $\lambda$ leaves $x$ unchanged (i.e. $x\lambda = x$). Note that any mnesor has $\tau$ as stabilizer [by (3)].

EXAMPLE. $\begin{bmatrix} France \\ Germany \end{bmatrix} IOC = \begin{bmatrix} France \\ Germany \end{bmatrix}$ where *IOC* stands for International Olympic Commitee.

Absorption property solutions are stabilizers.

PROOF: Let a mnesor $x$ be given. Since there exists a granular $\alpha$ such that $(x + y)\alpha = x$ [by (7)], we can substitute $(x + y)\alpha$ for $x$ in $x\alpha$:

$$\begin{aligned} x\alpha &= ((x + y)\alpha)\alpha \\ &= (x + y)(\alpha \otimes \alpha) \quad \text{[by (5)]} \\ &= (x + y)\alpha \quad \text{[by idempotence]} \\ &= x \end{aligned}$$

Note that $x = x\alpha = x\tau$, which shows that the general identity $x\lambda = x\mu$ basically may not be reduced to $\lambda = \mu$.

*Stabilizer sublattice*. The stablizers of $x$ make up a sublattice, written $\dfrac{\tau}{x}$.

PROOF: $x\lambda = x$ and $x\mu = x$. Thus, $x(\lambda \oplus \mu) = x\lambda + x\mu = x + x = x$. Thus, $\lambda \oplus \mu \in \dfrac{\tau}{x}$

$x(\lambda \otimes \mu) = (x\lambda)\mu = x\mu = x$. Thus $\lambda \otimes \mu \in \dfrac{\tau}{x}$

*Empty mnesor*. $e = 0$ iff $e\lambda = e$, for any granular $\lambda$.
That is, the stabilizer sublattice for $0$ is the whole lattice $L$.

PROOF: 
$$\begin{aligned} x + 0\lambda &= x + 0 + 0\lambda & \text{[by (1)]} \\ &= x + 0\tau + 0\lambda & \text{[by (3)]} \\ &= x + 0(\tau \oplus \lambda) & \text{[by (6)]} \\ &= x + 0\tau & \text{[by (2)]} \\ &= x + 0 & \text{[by (3)]} \\ &= x & \text{[by (1)]} \end{aligned}$$

That is, $x + 0\lambda = x$, for any mnesor $x$. Similarly, $0\lambda + x = x$. Thus, $0\lambda$ is the identity element itself.

And conversely, if $e\lambda = e$ holds for any granular, then $e\alpha = e$, where $(0 + e)\alpha = 0$ [ b y

absorption property]. Finally, $e\alpha = e$ and $e\alpha = 0$. Hence, $e = 0$

In case the lattice possesses a bottom element $\varepsilon$, then $x\varepsilon = 0$, for any mnesor $x$.

PROOF:  First, $(x\varepsilon)\lambda = x(\varepsilon \otimes \lambda) = x\varepsilon$, for any $\lambda \in L$
Thus, $x\varepsilon = 0$

*Annihilator*. A granular $\lambda$ is an annihilator for a given mnesor $x$ if it vanishes $x$ (i.e. $x\lambda = 0$).

EXAMPLE.  $\begin{bmatrix} India \\ Taiwan \end{bmatrix} EU = 0$

*Annihilator sublattice*. The annihilators of $x$ make a sublattice, written $\dfrac{\varepsilon}{x}$

PROOF: Let $\lambda, \mu \in \dfrac{\varepsilon}{x}$ be given.

Then, $x(\lambda \oplus \mu) = x\lambda + x\mu = 0$. Thus, $\lambda + \mu \in \dfrac{\varepsilon}{x}$

and $x(\lambda \otimes \mu) = (x\lambda)\mu = 0\mu = 0$. Thus, $\lambda \otimes \mu \in \dfrac{\varepsilon}{x}$